# AN ANALYSIS OF TWO COMMON REFERENCE POINTS FOR EEGS

*S. López, A. Gross, S. Yang, M. Golmohammadi, I. Obeid and J. Picone*

Neural Engineering Data Consortium, Temple University, Philadelphia, Pennsylvania, USA
{silvia.lopez, aaron.gross, scott.yang, meysam, obeid, picone}@temple.edu

*Abstract*— **Clinical electroencephalographic (EEG) data varies significantly depending on a number of operational conditions (e.g., the type and placement of electrodes, the type of electrical grounding used). This investigation explores the statistical differences present in two different referential montages: Linked Ear (LE) and Averaged Reference (AR). Each of these accounts for approximately 45% of the data in the TUH EEG Corpus. In this study, we explore the impact this variability has on machine learning performance. We compare the statistical properties of features generated using these two montages, and explore the impact of performance on our standard Hidden Markov Model (HMM) based classification system. We show that a system trained on LE data significantly outperforms one trained only on AR data (77.2% vs. 61.4%). We also demonstrate that performance of a system trained on both data sets is somewhat compromised (71.4% vs. 77.2%). A statistical analysis of the data suggests that mean, variance and channel normalization should be considered. However, cepstral mean subtraction failed to produce an improvement in performance, suggesting that the impact of these statistical differences is subtler.**

## I. Introduction

Diagnosis of clinical conditions such as epilepsy are dependent on electroencephalography (EEG), the recording of the brain's electrical activity through electrodes placed on the scalp, as shown in Figure 1. Delivering a conclusive diagnosis without an EEG is often unfeasible [1]. The key role played by this technique in the diagnosis of several neurological conditions coupled with the large amounts of time required by specialized neurologists to interpret these records, has created a workflow bottleneck – neurologists are overwhelmed with the amount of data that needs to be manually reviewed [2]. There is a great need for partial or complete automation of the EEG analysis process, and automated technology is slowly emerging to fill this void [3],[4]. The need for this data to be manually reviewed in real-time for clinical reasons further exacerbates the need for automatic interpretation technology.

Research has specifically focused on the task of ictal (seizure) EEG detection or identification. In [3], for instance, hidden Markov models (HMMs) are trained to recognize the ictal, interictal and postictal stages of the brain. The research presented in [5], on the other hand, describes a system that uses a wavelet-based sparse functional linear model with a 1-NN classifier for the classification of ictal EEGs. The same task was accomplished in [4] through the implementation of a Support Vector Machine (SVM) classifier. All these studies achieved detection accuracies in the range of 89% to 100%, even though clinical performance of commercial technology based on these approaches is significantly lacking [6][7].

Few studies, however, have addressed an important problem inherent to clinical recordings: the immense variability. All seizure detection studies previously cited limit the training and evaluation of their models to one or two homogeneous databases. The large variability among EEG channels and montages utilized in clinical EEGs is not usually taken into account for the generation and evaluation of the models. For example, in the TUH EEG Corpus [8], which is the basis for this study, there are over 40 different channel configurations and at least 4 different types of reference points used in the EEGs administered. It is unclear that whether this data can be modeled by a single statistical model, or whether special measures must be taken to account for this variability. Research fields such as speech recognition have dealt with this problem for many years using technologies such as speaker and channel adaptation [9], but these technologies have yet to be explored in EEG research.

The information yielded by an EEG channel is essentially the difference of electrical activity between two electrodes. In Figure 1, we show a typical EEG electrode pattern that includes common electrical reference points. Because changes in the electrode locations on the scalp present different electrical activity, the reference point used to measure a voltage has a significant impact on the nature of the voltage observed. In fact, since the conduction of these electrical signals through the brain is a highly nonlinear and noisy process, grounding plays a very important role in the quality of the observed signals.

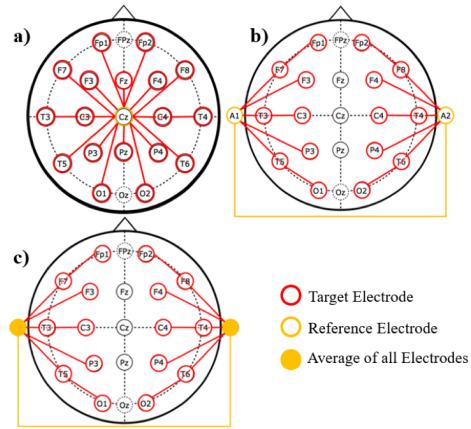

Figure 1. Three common referential montages are shown: a) the Common Vertex Reference ($C_z$), b) the Linked Ears Reference (LE) and c) the Average Reference (AR).

A differential view of the data, known as a montage, which consists of differencing the signals collected from two electrodes (e.g., Fp1-F7), is very common. In fact, neurologists are very particular about the type of montage used when interpreting an EEG. At Temple University Hospital (TUH), for example, a Temporal Central Parasagittal (TCP) montage [10] is very popular.

Of course, one might think that this problem is of little importance since most EEG analysis is done using differential voltages (e.g. Fp1-F7). In theory, the effects of a reference point would be cancelled via subtraction of two channels with the same reference point. In practice, the location of the reference point changes the nature of the waveforms considerably because the brain and scalp conduction paths are highly nonlinear [11].

The American Clinical Neurophysiology Society (ACNS) recognizes that there is a great variety of montages among EEG laboratories. Even though the ACNS has proposed guidelines for a minimum set of montages [10], several reference sites are still used depending on the purpose of the EEG recording [8]. Some commonly used reference schemes include:

- *Common Vertex Reference ($C_z$)*: uses an electrode in the middle of the head;
- *Linked Ears Reference (A1+A2, LE, RE)*: based on the assumption that sites like the ears and mastoid bone lack electrical activity, often implemented using only one ear;
- *The Average Reference (AR)*: uses the average of a finite number of electrodes as a reference.

The robustness of a state of the art machine learning system that decodes EEG signals depends highly on the ability of the system to maintain its performance with different variations of the data. The specific montage of a recording could potentially affect the operation of such systems in a negative way, which constitutes a fundamental problem, given the fact that EEG signals tend to present high variability in clinical settings [8].

This investigation will explore the statistical variations and effects that are produced by two different referential montages observed in the TUH EEG Corpus [8], LE and AR, on a machine learning system based on HMMs [12].

## II. EXPERIMENTAL DESIGN

The TUH EEG Corpus **Error! Reference source not found.** is the largest, publicly available source of clinical data in the world. The referencing systems that are compared in this study are the ones that predominate in this corpus: Linked Ears Reference (LE) and Averaged Reference (AR) and (43.8% and 46.5% of the data respectively). The large amounts of data available in TUH EEG (approximately 16,500 files each), was the main motivation for the selection of these particular referential systems.

The study of the two referential systems was divided into three types of analyses: (1) simple descriptive statistics, (2) analysis of variance using Principal Component Analysis (PCA) [13] and (3) a comparison of the performance obtained from our standard HMM baseline system that uses models trained separately for each class.

Feature extraction for EEG signals was performed using a standard approach described in [12] and shown in Figure 2. The frame and window durations for feature extraction are 0.1 and 0.2 seconds respectively. The base features were used in a calculation that produced their first and second derivatives. It is important to note that the second derivative was not calculated for the differential energy feature, because it was proven to be redundant in previous studies [12].

The final feature vector that was used as an input for the experiments had a dimension of 26, with 9 of those features being the base, or absolute, features, and the rest being derivatives of the original features. The number of features used was varied depending on the experiment. In some experiments, only absolute features (9 features) were used because these are more appropriate for studying basic statistical properties since they map directly to spectral characteristics of the signal.

The descriptive statistics of the data were calculated through a simple computation of the mean and variance for each class (LE and AR). The global mean and variance for all the data were also calculated in order to determine the significance and direction of the bias. For this particular part of the study, 16,840 LE files and 17,858 AR files were used, meaning that 48.5% of the data were referenced to LE while 51.5% of the data was referenced to AR. Note that for this part of the study, only the base features were used.

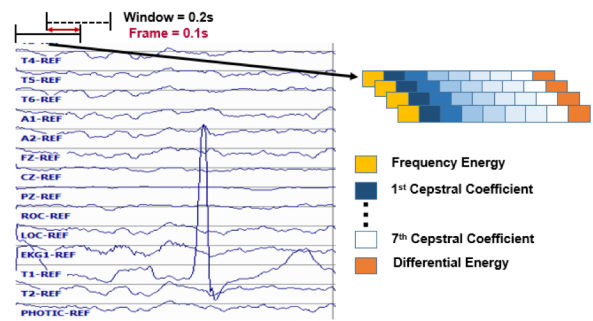

Figure 2. Base features calculated through a cepstral coefficient-based approach that uses frame and window durations of 0.1 and 0.2 seconds respectively.

Following the descriptive statistical analysis, a PCA was performed on the features. The 9-dimensional mean vector, μ, and covariance matrix Σ of the data were computed, which was followed by the calculation of the eigenvalues and vectors of the covariance matrix. The eigenvalues and eigenvectors were then analyzed and compared to the comparable eigenvector in the opposite class. This was done to assess the importance of each component in the overall weighting of the feature vector.

A machine learning system was used to evaluate the mismatch between feature vectors from each class. This baseline system was a traditional HMM system described in [12]. This specific system, however, was trained to detect two different types of events: (1) seizures (SEIZ), and (2) background (BCKG). To assess the mismatch between feature vectors, we trained statistical models with only LE features (LE model), only AR features (AR model) and the combination of both types of features combined (LE+AR model). The models were evaluated in similarly divided evaluation sets (LE data only, AR data only and LE+AR data). The training sets were comprised of 44 EEG records for each class (LE and AR) and the evaluation set had 10 EEG records per class. All of the records in both the training and evaluation sets came from unique patients, which implies that 108 patients were represented in the total dataset.

Speech recognition systems have been generally successful in mitigating the influence of channel variations. Feature normalization techniques, such as Cepstral Mean Normalization (CMN) [14], are well-established techniques that enhance the robustness of these systems. We also report on a pilot experiment using CMS to offset any biases between montages.

### III. RESULTS AND DISCUSSION

Descriptive statistics were calculated for both classes per feature type as an initial analysis. Table 1 presents a summary of the findings. These statistics demonstrate that there is a great variation in the means and variances for each base feature, indicating that the characteristics that describe these two sets are very different in the frequency domain. We also examined individual channels and observed a comparable amount of variation.

PCA analysis provides a more complete analysis of the differences between montages. The percent variance explained by each eigenvalue is presented in Figure 4 for each of the montages. Figure 6 compares the eigenvectors. We observe that the first PCA component explains a much higher portion of the variance for the LE data than for the AR data. This analysis was supported in Figure 6. The eigenvectors show similar behavior in the energy features and the lower cepstral coefficients. The lower order eigenvectors, which correspond to large eigenvalues, weight the higher cepstral coefficients more heavily. These features, whose eigenvectors show opposite polarity, correspond to beta waves (13 Hz – 30 Hz) frequently present in normal recordings.

The recognition experiments on seizure detection were much more revealing. A Detection Error Tradeoff (DET) curve for each of these experiments is presented in Figure 7 while the detection rate is summarized in Table 2. Best performance is obtained by training on the entire dataset (LE+AR) and evaluating only on LE. However, the performance of this model on the AR data set is degraded, causing the overall performance on the combined data to suffer. The AR model is the one with the least amount of variability when tested on different evaluation sets.

The results presented in

| Train/Eval | LE | AR | LE+AR |
|---|---|---|---|
| LE | 77.19% | 72.89% | 78.52% |
| AR | 55.92% | 61.41% | 60.89% |
| LE+AR | 68.60% | 68.25% | 71.40% |

Table 2 support the fact that the three models, AR, LE and LE+AR are fundamentally different. The bias between the montages that can be seen in

|  | Mean | | Variance | |
|---|---|---|---|---|
| Feature | LE | AR | LE | AR |
| Ef | 1.685 | 12.390 | 49.560 | 19.368 |
| c1 | 2.296 | 1.949 | 0.891 | 1.171 |
| c2 | 0.991 | 0.677 | 0.510 | 0.675 |
| c3 | 0.320 | 0.296 | 0.166 | 0.250 |
| c4 | -0.060 | -0.009 | 0.107 | 0.128 |
| c5 | -0.026 | 0.037 | 0.037 | 0.050 |
| c6 | -0.007 | -0.035 | 0.024 | 0.027 |
| c7 | 0.045 | 0.042 | 0.017 | 0.016 |
| Ed | 1.887 | 3.001 | 23.298 | 21.824 |

Table 1. Summary of the descriptive statistics of the elements of the feature vector by montage.

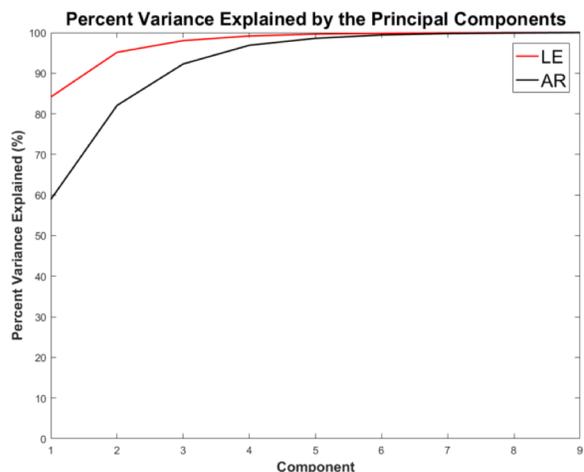

Figure 3. Percent variance explained by each principal component for each referential montage type.

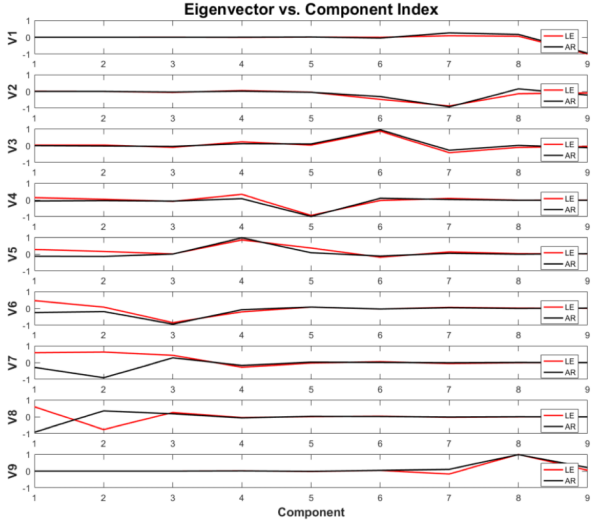

Figure 4. The amplitudes of the eigenvectors for each montage are shown. Note that components 2-8 represent the cepstral features, component 1 represents frequency domain energy and component 9 represents differential energy.

|  | Mean | | Variance | |
| --- | --- | --- | --- | --- |
| Feature | LE | AR | LE | AR |
| Ef | 1.685 | 12.390 | 49.560 | 19.368 |
| c1 | 2.296 | 1.949 | 0.891 | 1.171 |
| c2 | 0.991 | 0.677 | 0.510 | 0.675 |
| c3 | 0.320 | 0.296 | 0.166 | 0.250 |
| c4 | -0.060 | -0.009 | 0.107 | 0.128 |
| c5 | -0.026 | 0.037 | 0.037 | 0.050 |
| c6 | -0.007 | -0.035 | 0.024 | 0.027 |
| c7 | 0.045 | 0.042 | 0.017 | 0.016 |
| Ed | 1.887 | 3.001 | 23.298 | 21.824 |

Table 1 was addressed through the implementation of CMN, in the hopes of stabilizing the systems. Unfortunately, CMN did not prove to be as successful with EEG data. Figure 5 shows that the performance with CMN is worse for all cases except the AR model evaluated on LE data.

## IV. SUMMARY

EEG machine learning technology should be robust to any type of EEG signal. The ability to train channel-independent models, or to maintain performance across different montages, is extremely important in clinical settings, where there is not one specific standard way to conduct the recordings. Our analysis of the two different referential montages that represent the majority of the data in the TUH EEG Corpus, Linked Ears Reference (LE) and Averaged Reference (AR), shows that there are systematic differences in the statistics of the data. Though our existing baseline system is capable of addressing these variations, it seems likely that some form of channel normalization should improve performance and reduce the variance of the model.

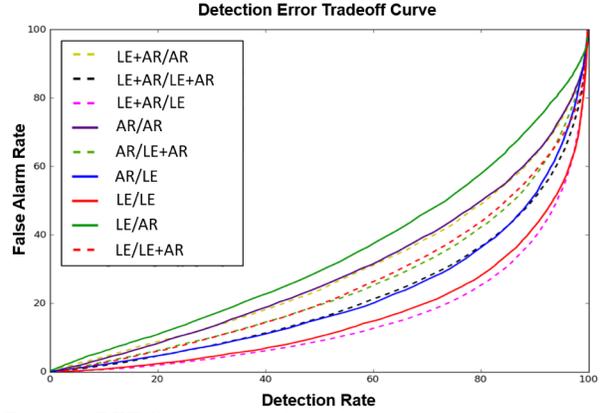

Figure 6. DET Curves for each of the recognition experiments. The first montage indicator refers to the data used for training, while the second one refers to the evaluation set. For example, LE+AR/AR refers to a model trained with LE+AR data and evaluated with AR data.

Cepstral mean normalization (CMN) was implemented in order to address the mean bias that is present in the two different referential systems. Our results indicate that this technique was not as successful in the EEG domain as it was in speech. Additional investigation into this topic is warranted. This paper has shown that finding and implementing a successful normalization approach for clinical EEGs would allow the data to be mixed, thereby making the overall corpus more useful for machine learning research.

ACKNOWLEDGEMENTS

Research reported in this publication was supported by the National Human Genome Research Institute of the National Institutes of Health under award number U01HG008468. The content is solely the responsibility of the authors and does not necessarily represent the official views of the National Institutes of Health.

This material is also based in part upon work supported by the National Science Foundation under Grant No. IIP-

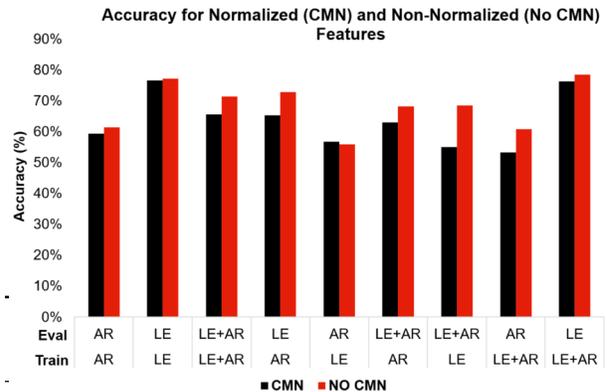

Figure 5. Performance comparison for the normalized and non-normalized systems.

1622765. Any opinions, findings, and conclusions or recommendations expressed in this material are those of the author(s) and do not necessarily reflect the views of the National Science Foundation. The TUH EEG Corpus work was funded by (1) the Defense Advanced Research Projects Agency (DARPA) MTO under the auspices of Dr. Doug Weber through the Contract No. D13AP00065, (2) Temple University's College of Engineering and (3) Temple University's Office of the Senior Vice-Provost for Research.